\documentclass[12pt]{article}
\pdfoutput=1

\usepackage{amsmath,amssymb,amsfonts}
\usepackage{cite}

\usepackage{geometry}                
\usepackage{graphicx}
\usepackage{enumerate} 

\usepackage[usenames,dvipsnames]{xcolor}
\definecolor{linkCol}{rgb}{0.0,0.19,0.59} 

\usepackage[pdftex, 
colorlinks,
citecolor=linkCol,
pdfauthor=Sergio,
]{hyperref}

\title{Correlation function for the Grid-Poisson Euclidean matching  on a line and on a circle}
\author{  
Elena Boniolo~\thanks{Actual address: School of Physics,
HH Wills Physics Laboratory,
University of Bristol,
Tyndall Avenue,
Bristol,
BS8 1TL,
United Kingdom} \\[0.3cm]
   {\small\it Dipartimento di Fisica dell'Universit\`a degli Studi di Milano}\\
  {\small\it via Celoria 16, I-20133 Milano, ITALY}
  \\
  {\small\tt elena.boniolo@gmail.com
} \\[0.3cm]
Sergio Caracciolo
\\[0.3cm]
  {\small\it Dipartimento di Fisica dell'Universit\`a degli Studi di Milano, and INFN,}\\
  {\small\it via Celoria 16, I-20133 Milano, ITALY}\\
  {\small\tt Serg{}io.Caracci{}olo@mi.i{}nfn.it
  }\\[0.3cm]
Andrea Sportiello
\\[0.3cm]
  {\small\it LIPN, and CNRS, Universit\'e Paris 13, Sorbonne Paris Cit\'e, }\\
  {\small\it 99 Av. J.-B. Cl\'ement, 93430 Villetaneuse, FRANCE}\\
  {\small\tt An{}drea.Spor{}tiello@lipn.fr
}}

\newcommand{\be}{\begin{equation}}
\newcommand{\ee}{\end{equation}}

\newcommand{\bulB}{{\bullet}}
\newcommand{\bulW}{{\circ}}
\newcommand{\bulX}{{\,\cdot\,}}

\newcommand{\ef}[1]{\, #1}

\def\reff#1{(\protect\ref{#1})}

\def\var{\operatorname{var}}
\def\cov{\operatorname{cov}}
\def\sgn{\operatorname{sgn}}

\newtheorem{defin}{Definition}[section]

\newtheorem{prop}[defin]{Proposition}

\def\prf{\par\noindent{\bf Proof.\enspace}\rm}
\newcommand{\qed}{\quad $\Box$ \medskip \medskip}
\newcommand{\espo}{p}

\begin{document}

\maketitle

\begin{abstract}
We compute the two-point correlation function for spin configurations
which are obtained by solving the Euclidean matching problem, for one
family of points on a grid, and the second family chosen uniformly at
random, when the cost depends on a power $p$ of the Euclidean
distance. We provide the analytic solution in the thermodynamic limit,
in a number of cases ($p>1$ open b.c.\ and $p=2$ periodic b.c., both at criticality), and
analyse numerically other parts of the phase diagram.
\end{abstract}

\clearpage

\section{Introduction}

\subsection{The problem}

The \emph{Matching Problem} is an optimisation problem, in which the
set of feasible configurations consists of possible maximal matchings
in a bipartite graph, and the cost function is the sum of the costs on
the individual chosen edges.

Let us call ${\cal K}_{N,M}$ the complete bipartite graph, so that
the set of vertices $V$ is partitioned in 
$V = {\cal R} \cup {\cal B}$, where ${\cal R} = \{r_1, \dots, r_N\}$
is the set of {\em red} vertices, and ${\cal B}=\{b_1, \dots, b_M\}$
is the set of {\em blue} vertices.

Assume, without loss of generality at this point, that $N \leq
M$. Maximal matchings are thus the set 
$\Pi$
of injective mappings $\{\pi: {\cal R} \to {\cal B}\}$.
Given a collection of \emph{weights} $w(r_i, b_j) =
w(i,j) \in {\mathbb R}^+ \cup \{+\infty\}$,
we assign to each maximal matching $\pi \in \Pi$ the cost $E(\pi)$
defined as
\be
E(\pi) = 
\sum_{i=1}^N w(i, \pi(i))
\ef.
\label{eq.defcost}
\ee
The {\em optimal} matching $\pi_{\rm opt}$, and optimal cost
$E_{\rm opt}$, are the quantities
that realize the minimum cost
\be
E_{\rm opt} = E(\pi_{\rm opt}) = \min_{\pi \in \Pi} E(\pi) \, .
\ee
This problem models a variety of concrete applications in Optimisation
Theory. In particular, it is the discrete version of a problem
introduced by Monge~\cite{monge}, back in 1781, where blue and red
sites corresponded to production and exploitation sites of some
resource, and the matching is optimising the cost of
transportation. The continuous version of the problem, in which one
has to find the mapping which minimises the transport condition
between two given measures (the red and the blue ones) is also of
relevance and is studied under the name of Monge--Kantorovi\v{c}
problem~\cite{Villani}.

The research of the optimal matching in a bipartite weighted graph is
usually called the {\em Assignment Problem}. Both the bipartite and
non-bipartite optimal matching problems are of polynomial complexity,
and the bipartite case has a considerably simpler algorithm.

A classical polynomial algorithm for the Assignment Problem
is due to Kuhn~\cite{kuhn}, who called it {\em Hungarian Algorithm}
as a tribute to the country of origin of the authors of the two main
lemmas on which it is based, K\H{o}nig and Egerv\'{a}ry.  As reported
by Knuth~\cite{knuth:sgb}, after the work of Munkres~\cite{munkres}
for speeding up a certain `recovering procedure', the complexity is
cubic in~$M$.

Questions of statistical nature arise when the set of weights are
stochastic variables, and the optimal quantities $\pi_{\rm opt}$ and
$E_{\rm opt}$ are analysed probabilistically.  In particular, we are
motivated to consider these problems because of the close connection
between random optimization problems and the statistical mechanics of
disordered 
systems~\cite{mezard, hartmann, percus, bouchaud, montanari}. Indeed,
when the weights $w (i, j)$ are equally distributed independent random
variables, drawn from a large range of single-weight distributions, by
using the celebrated {\em replica trick}, Mez\'ard and Parisi could
compute the average cost for both the matching on the complete
graph~\cite{MP1}, the {\em random matching problem}, and on the
complete bipartite graph~\cite{MP2}, the 
{\em random bipartite matching problem}.  See~\cite{aldous} for a
derivation without replicas.

\subsection{Considerations at generic dimension $d$}

The fact that the weights are equally distributed implies, in
particular, that the stochastic problem has no underlying
finite-dimensional geometry, i.e.\ it is a \emph{spherical}, or
\emph{mean-field} disordered problem.
The implementation of the model we are going to discuss, named the
{\em Grid-Poisson matching problem}, is instead naturally embedded in
a finite dimension~$d$.

In this paper we will be mainly concerned with the (much simpler) case
$d=1$. However, in this introduction we supply a number of
observations in the case of generic~$d$.

For $L$ an integer, define the box $\Lambda = [0, L]^d \subset
{\mathbb R}^d$.  The set of red vertices $\cal R$ is chosen to be
${\cal R} = \Lambda \cap ({\mathbb Z}+\frac{1}{2})^d$, i.e.\ the set of
$N=L^d$ points within $\Lambda$ that have all semi-integer
coordinates.  The set of blue vertices, $\cal B$, is a set of $M$
points chosen uniformly at random in $\Lambda$.
Introducing a further parameter $\espo > 0$,
the weight is taken to be the corresponding power of the Euclidean
distance $d(i, j)$ between $r_i$ and $b_j$
\be
w (i,j) = \big( d(i,j) \big)^\espo
\label{pesi}
\ee
In such a case, we say that we have 
\emph{open boundary conditions}. In the variant in which $\Lambda$ is
compactified on a torus, and $d(i,j)$ is the minimal distance among
the translation images,\footnote{I.e., in such a case,
  $d\big((x_1,\ldots,x_d),(y_1,\ldots,y_d)\big) = \min_{\nu \in
    \mathbb{Z}^d} \sqrt{\sum_{a=1}^d (y_a-x_a-\nu_a L)^2}$.}
we say that we have 
\emph{periodic boundary conditions}.

The case of main interest is $\espo=1$, where the total cost has a
direct pictorial interpretation as the total length of the
segments. It is also the value at which the function $x \to x^\espo$
changes its behaviour (it is concave for $0<\espo<1$, and convex for
$p>1$), a property of relevance for the problem at hands. The cases of
$\espo$ an even integer are also special, as the Euclidean distance to
the power $\espo$ can be expressed as a polynomial in the Cartesian
coordinates of the points.

A related model, the {\em Poisson-Poisson matching problem}, where
both the red and blue sets occur as independent Poisson processes of
equal intensity, has been considered in~\cite{akt, holroyd,clps}. The
{\em Euclidean monopartite matching problem} has also been studied by
using the replica trick~\cite{MP2}, by taking corrections to the
random matching problem.  The {\em minimax Grid-Poisson matching
  problem} has been studied in~\cite{leighton, shor, sicuro}. In this problem,
the cost function is changed from (\ref{eq.defcost}) into
\be
E^{\rm m.m.}(\pi) = 
\max_{1 \leq i \leq N} w(i, \pi(i))
\ef.
\label{eq.defcostMM}
\ee
It is easy to see that this problem is a limit $p \to +\infty$ of the
class of problems defined above, namely, for the same set of points,
$\pi_{\rm opt}^{\rm m.m.} = \lim_{p \to \infty} \pi_{\rm opt}^{(p)}$
and
$E_{\rm opt}^{\rm m.m.} = \lim_{p \to \infty} 
\big( E_{\rm opt}^{(p)} \big)^{\frac{1}{p}}$.

In a generic configuration of points, each red point $r_i$ has a
unique nearest blue point $b_{j(i)}$, and vice versa. If $j(i) \neq
j(i')$ for all $i \neq i'$, the corresponding matching is easily
certified to be optimal, for all values of $p$ simultaneously.  Such a
simple situation occurs with increasing probability for $M/N \to
\infty$, and, in a symmetric way exchanging red and blue, for $M/N \to
0$. Conversely, when $N \sim M$ we expect competing effects for the
colliding pairs $(i,i')$ such that $j(i) = j(i')$, analogous to
frustration in disordered systems, and long-range correlations, of the
order of the size of the system, may arise.  For this reason, we shall
look at the {\em continuum limit}, in which both $N$ and $M$ become
infinitely large, by keeping fixed the ratio
\be
\rho := \frac{M}{N}\, .
\ee
Let us say that $\xi=\xi(\rho)$ is a scale of correlations in the
system, at given density, and in the continuum limit.  As it will turn
out, $\xi$ diverges at only one \emph{critical} value of the density
$\rho^*$, like in a second-order phase transition.  Under this
assumption, it is clear that, in the Poisson-Poisson problem,
$\rho^*=1$. This seems to remain numerically true, although
theoretically more subtle, in the Grid-Poisson case.

An argument for justifying both the second-order character of the
behaviour of $\xi(\rho)$, and the stability of the position of the
critical density at the self-dual value, is through a coarse-grain
analysis.

More generally, consider the case in which red and blue points arise
from two independent processes, one of which being Poisson, the other
one having fluctuations at most as in the Poisson case (it may be
another Poisson process, or a regular lattice, or a Determinantal
process, \ldots).  Consider boxes in $\Lambda$ of size
$X=v^{\frac{1}{d}}$, and assume that $1 \ll X \ll L$. The average
number of red and blue points are $v$ and $\rho v$, respectively. The
fluctuations on these numbers are of the order of $\sqrt{v}$. As soon
as $\sqrt{v} > \frac{1}{|\rho-1|}$, with large probability there are
enough blue points within each box to be matched to the red
points. Thus, the matching obtained by solving the problem separately
in each box is maximal, and has all edges of length bounded by $\sim
v^{\frac{1}{d}}$. Conversely, if $\rho=1$, at all coarse-grain scales
we will have important fluctuations. A first coarse-graining at scale
$v^{\frac{1}{d}}$ will leave $\sim \sqrt{v}$ unmatched red or blue
points per box, independently on each box. A further coarse-graining
at scale $(k v)^{\frac{1}{d}}$, for the remaining points, gives around
$k v^{\frac{1}{2}}$ points of each colour, and around $k^{\frac{1}{2}}
v^{\frac{1}{4}}$ excess of points of one colour, thus a feature
analogous to a single coarse-graining at scale 
$(k \sqrt{v})^{\frac{1}{d}}$. The self-similarity of the
coarse-graining procedure is a signature of an interesting behaviour
under the group of renormalisation, and of long-range correlations.

For these reasons, we predict that, also in the Grid-Poisson case the
critical density is for $\rho = 1$, and set
\be
t := \rho - 1
\ee
as a useful shortcut for the {\em reduced temperature}, which vanishes
at the critical point.

Our aim is to study the correlations which emerge in the 
{\em scaling region} around the critical point.

\subsection{Observables}

As we have chosen to sample red and blue points through two distinct
procedures, we no longer have a
symmetry of the problem under
exchange of red and blue points. Thus in the case $N \leq M$ the
maximal matchings are injections from $\mathcal{R}$ to $\mathcal{B}$,
and in the case $N \geq M$ the
maximal matchings are injections from $\mathcal{B}$ to
$\mathcal{R}$. 
(Of course, for $M=N$ we have bijections, i.e.\ permutations.)

Let $r_i$, with $i=1, \dots, N$ be the vector of integer coordinates
for the red points and $b_j$, with $j=1, \dots, M$, the vector of real
coordinates for the blue points.  We define a collection of
$\min(N,M)$ vectors $\varphi_i$ associated to the red points covered
by the optimal matching. If $(i,j)$ is an edge in $\pi_{\rm opt}$, we
set
\be
\varphi_i := b_j - r_i
\ee
In particular,
\be
E_{\rm opt} = \sum_{i =1}^{\min (N,M)} |\varphi_i|^\espo 
\ef.
\ee
The quantities $\varphi_i$ behave as $O(d)$ vectors on the lattice. As
for $O(n)$ models in (non-disordered) statistical mechanics systems,
we expect that the spontaneous symmetry breaking, if any, occurs in
the angular degrees of freedom. For this reason we also introduce, for
the same set of indices $i$,
a \emph{spin variable} $\sigma_i$
%
%
\be
\sigma_i  := \frac{\varphi_i}{|\varphi_i|} 
\ee
Let $\sigma_i=0$ if $M<N$, and $i$ is not covered in the optimal
matching.

Thus, a model, here given by a triple $(d,L,\espo)$, induces a measure
$\mu(\sigma)$ over the corresponding set of spin variables.
We shall characterize the critical behaviour of our model by looking
at the correlation function
\be
G(x,y) := 
\sum_\sigma \mu(\sigma)\; \sigma_x \cdot \sigma_y = \langle \sigma_x
\cdot \sigma_y \rangle\, 
\ee
with $x$, $y$ points of our grid.

At this point, an advantage of the Grid-Poisson version of the
problem, w.r.t.\ the Poisson-Poisson version, becomes evident. The
statistical properties of $G(x,y)$ are more easily investigated
numerically in the first case, in particular for periodic boundary
conditions, where $G(x,y)=G(y-x)$ takes as argument an integer-valued
vector in $\mathbb{Z}^d$, instead of a real-valued vector.


Define the correlation function averaged over pairs of points with the
same distance
\be
G(r; L, t) = \frac{\sum_{(x,y): d(x,y) = r} G(x, y)}
{\sum_{(x,y): d(x,y) = r} 1}
\ef.
\label{def:G}
\ee
In the region near criticality we expect {\em finite-size scaling} of
the correlation functions. In particular for the two-point function
in~\reff{def:G}, this means that it must be a homogeneous function of
its arguments, according to
\be
G(r; L, t) \, = \, L^\alpha\, 
F\left( \frac {r}{L}, t\,L^\frac{1}{\nu} \right) 
\label{scaling}
\ee
where the exponents $\alpha$ and $\nu$ and the function $F$ are 
{\em universal}, that is are common to other models in the same
universality class.  In particular, for our family of models, we
expect them to depend on the dimensionality $d$, and the exponent
$\espo$ by which the Euclidean distance enters the cost function.
In this paper they will be the main argument of our interest.

We repeat that in this paper we study the one-dimensional version of
the problem, in the two variants of boundary conditions, open and
periodic.

\section{Open boundary conditions}

In this section we analyse the one dimensional Matching Problem with
open boundary conditions.

\subsection{Properties of the optimal matching}

We start from analysing some general properties of the problem on a
line with open boundary conditions.

We first discuss the consequences of the choice of the exponent $\espo$
which appears in the weights~\reff{pesi}.
Let us compare the cost of matchings such that two given red points
$r_1$, $r_2$ are matched to two given blue points
$b_1$, $b_2$ (in one of the two orders), given that the rest of the
configuration is the same.  If we determine which of the two orders is
the best, we have a criterion for excluding that the other ordering is
part of the optimal matching.  The analysis goes through a case study,
for the $4!=24$ possible orderings of 
$\{ r_1, r_2, b_1, b_2\}$ along the line. Of course, the discrete
symmetries reduce the analysis to only three cases, that we denote by
the pictograms
$[\bulB \bulB \bulW \bulW]$, 
$[\bulB \bulW \bulB \bulW]$ and
$[\bulB \bulW \bulW \bulB]$.
%


Let $T_1$ be the cost of the matching in which the leftmost red point
goes with the leftmost blue one, and $T_2$ the cost of the other
possible matching.  We shall call the first case {\em ordered}.
More generally, we shall call {\em ordered} a matching such that, for
all pairs of edges $(r_1,b_1)$, $(r_2,b_2)$, if $r_1<r_2$ then 
$b_1<b_2$.
If we draw a matching
with arcs on the upper half-plane, some of the arcs may cross. We call
a matching \emph{crossing} if this occurs, and \emph{non-crossing} otherwise.

\begin{figure}[t]
 \centering
\begin{center}
\includegraphics[width=0.80\textwidth]{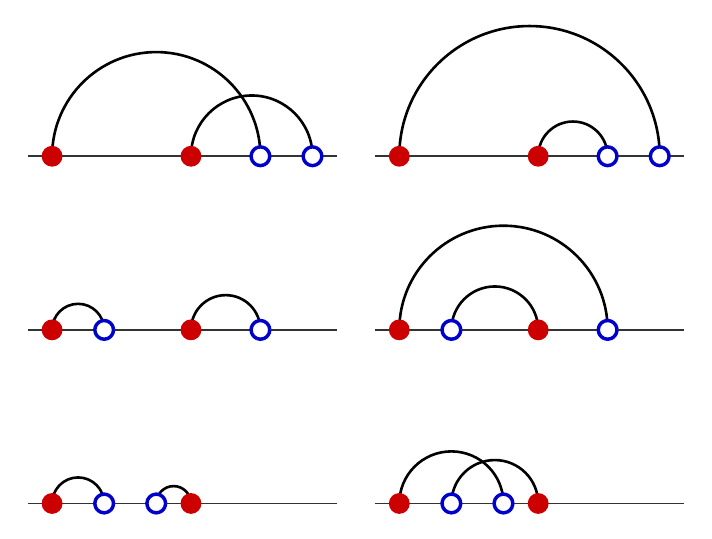}
\end{center}
\caption{Matchings for size-2 instances. Each line corresponds to a particular disposition of the blue points. The matchings on the left are all {\em ordered}. The first and the last matching are {\em crossing}.}
\label{fig:ambiguity1}
\end{figure}

{\bf First case, $[\bulB \bulB \bulW \bulW]$.}
Let the positions be, from left to right, 
$z, z+y, z+ y+ x_1, z+ y +x_2$ with $x_2>x_1$ (see
Fig.~\ref{fig:ambiguity1}).  As the distances are invariant under
translations we can choose $z=0$ and we can also set $y=1$, by
choosing the unit of lengths.

The first matching is ordered, the second one is non-crossing.
The costs of the two matchings are
\begin{align}
T_1 = & \,  (1+x_1)^\espo + x_2^\espo \\
T_2 = & \,  (1+x_2)^\espo + x_1^\espo \ .
\end{align}
Now, $T_1 \leq T_2$ if and only if 
\be
(1+x_1)^\espo - x_1^\espo \leq (1+x_2)^\espo - x_2^\espo 
\ee
The function $f(x) = (1+x)^\espo - x^\espo$ is always increasing,
respectively decreasing, on $\mathbb{R}^+$, when $\espo > 1$,
respectively $\espo <1$ (and is $f(x)=1$ for $\espo=1$).

Thus, for $\espo >1$ the ordered matching has a lower cost.  For
$\espo =1$ the two matchings have the same cost. For $\espo <1$ the non-crossing
matching has lower cost.

{\bf Second case, $[\bulB \bulW \bulB \bulW]$.}
Similarly to the case above, by translating and scaling we can get rid
of two parameters.
Let the positions be $0, 1-x_1, 1, 1+x_2$ with $0 < x_1 <1$.
In this case, the two possible matchings are both non-crossing.
The costs are now
\begin{align}
T_1 = & \,  (1-x_1)^\espo + x_2^\espo \\
T_2 = & \,  (1+x_2)^\espo + x_1^\espo \ .
\end{align}
Now, $T_1 \leq T_2$ if and only if
\be
(1-x_1)^\espo - x_1^\espo \leq (1+x_2)^\espo - x_2^\espo \, .
\ee
For $\espo \geq 1$ this is always true, because
\be
(1-x_1)^\espo - x_1^\espo \leq 1 \leq (1+x_2)^\espo - x_2^\espo \, 
\ee
which implies that the ordered matching has a lower cost.

For $\espo <1$, at fixed $x_2$, the ordered matching has a lower cost
only for a range of values $x_1$.  For example, let $\espo = 1/2$ and
set $z= \sqrt{1+x_2} - \sqrt{x_2}$, which maps the domain $x_2 \in
[1,\infty)$ into $z \in (0,1]$.  Then the ordered matching has a lower
cost if and only if
\be
x_1 >  \frac{1}{2} \, \left( 1 - \sqrt{2 z^2 - z^4} \right)
\ee
which does \emph{not} cover the full domain $x_1 \in [0,1]$, in
general. 


{\bf Third case, $[\bulB \bulW \bulW \bulB]$.}
Let the positions be $0, x_1, x_2, 1$ with $0 < x_1 < x_2< 1$.
In this case the ordered matching is non-crossing, the other one is crossing.
The costs are now
\begin{align}
T_1 = & \,  x_1^\espo +  (1-x_2)^\espo  \\
T_2 = & \,  x_2^\espo +  (1-x_1)^\espo \ .
\end{align}
Now, $T_1 \leq T_2$ if and only if
\be
  x_1^\espo - (1-x_1)^\espo \leq x_2^\espo - (1-x_2)^\espo 
\ee
which is always the case for $\espo \geq 0$ because the function
$f(x) = x^\espo - (1-x)^\espo$ is increasing on $[0,1]$.

From this analysis we deduce the following two statements.

\begin{prop}
\label{prop.sec2}
For $\espo > 1$ the optimal matching is ordered.
For $\espo < 1$ the optimal matching is non-crossing.
For $\espo = 1$ there exists an optimal matching which is ordered, and
one which is non-crossing.
\end{prop}

\prf Assume by absurd that the optimal matching is not ordered, and
consider a pair of edges $(r_1,b_1)$, $(r_2,b_2)$ which certify this.
From the previous analysis we see that by re-ordering them we obtain a
cost which is strictly lower (resp.\ weakly lower) for $\espo > 1$
(resp.\ for $\espo = 1$). The argument for the non-crossing statement
is analogous.\qed

Let us remark that when $N=M$, that is where we expect criticality,
and when $\espo>1$, the solution of the matching problem is very
simple because there is only one ordered matching.\footnote{The
  analogous statement, for $\espo<1$ and non-crossing matchings, does
  not hold because the pattern $[\bulB \bulW \bulB \bulW]$ has two
  potentially good non-crossing pairings.}
Since we are on a line, we can label red and blue points in increasing
order.  The solution of the matching problem is the one in which the
$i$-th red point is associated with the $i$-th blue point.  Then
$
\varphi_i  = b_i - r_i
$
and
$
\sigma_i = \sgn \varphi_i \in \{ -1, 1 \}
$
is the Ising spin variable that we can associate with the
solution. 
Furthermore, as $\pi_{\rm opt}$ does not
change in the full range $\espo>1$, all the geometric quantities (and
in particular the variables $\sigma_i$) are studied simultaneously for
all $\espo$ in this range.  For definiteness, we shall take $\espo=2$
as our preferential case in this range, as, in the case of periodic
boundary conditions, for this value we have an important
simplification.

When $\espo = 1$, almost surely on any optimal matching there is a
finite fraction of pairs of edges, with neighbouring red indices, in
the pattern $[\bulB \bulB \bulW \bulW]$. This suggests that almost
surely there is a large degeneracy of the optimal configuration.

\subsection{Numerical results}

We begin from the case $\espo=2$.

\begin{figure}[t!]
\begin{center}
\includegraphics[width=0.90\textwidth]{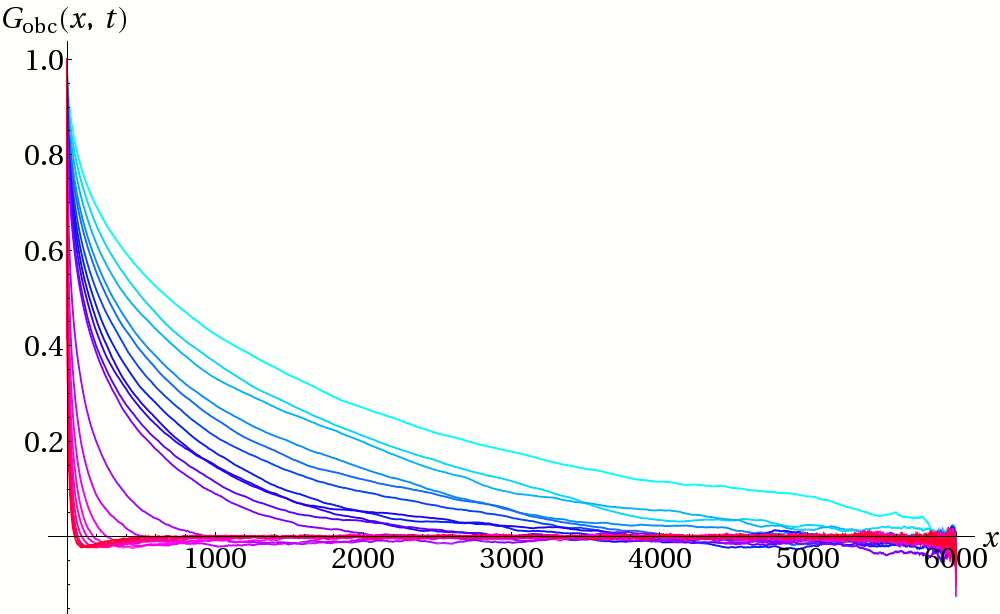}
\end{center}
\caption[]{The correlation function with open boundary conditions at size 6000. 
From cyan to purple, $t= $ 0, 0.001, 0.002, 0.003, 0.004, 0.005, 0.006, 0.007, 0.008, 0.009, 0.01, 0.02, 0.03, 0.04, 0.05, 0.06, 0.07, 0.08, 0.09, 0.1. }
\label{fig:dim1fuoriPcPiatto}
\end{figure}

In Fig.~\ref{fig:dim1fuoriPcPiatto} we report the correlation function
$G(r; L, t)$ for various choices of the parameter $t$ when the size of
the system is $L=6000$.  Each curve is the mean over $10^3$ instances
for the positions of the blue points.  Of course, The shape of the
functions for the corresponding negative values of $t$ are
undistinguishable.

We notice that $G(r; L, t)$ presents two ranges of behaviour. If
$|t|<\bar{t}(L)$  (with $\bar{t}(L)\approx 0.01$ when $L=6000$) the
function is strictly positive, it is decreasing with $r$, and has,
therefore,  a minimum at $r=L$. 
On the other hand, if $|t|>\bar{t}(L)$, the shape is different. It
reaches a minimum at an intermediate value $r'$, then goes up again
approaching zero as $r\to L$.  Whenever finite-size scaling holds for
the two-point function~\reff{scaling}, for large $L$, we get that
$\bar{t}(L) L^{\frac{1}{\nu}}$ is constant.

\begin{figure}[t!]
 \begin{center}
\includegraphics[width=0.90\textwidth]{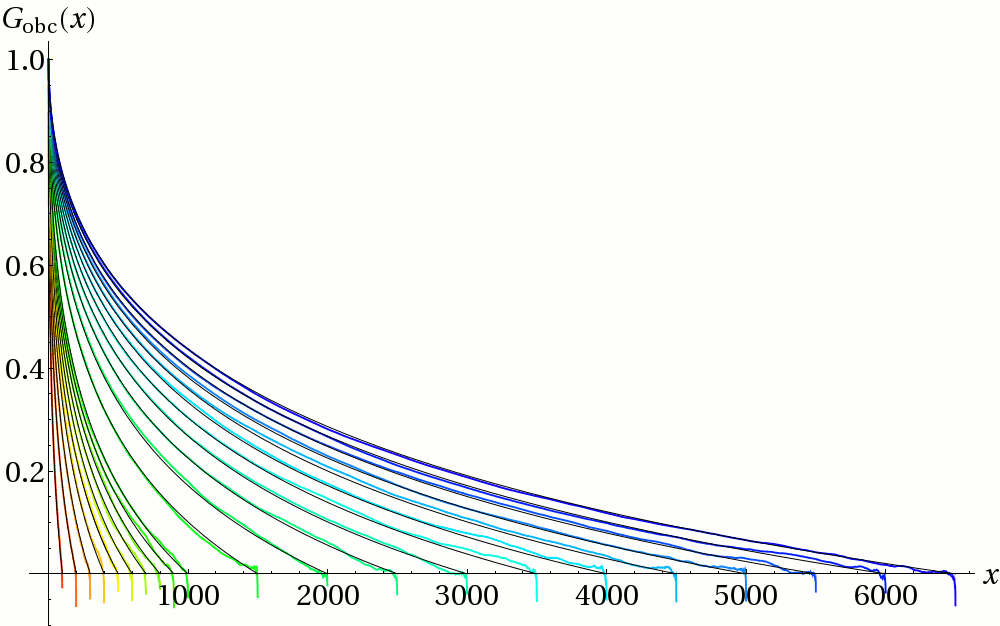}
\includegraphics[width=0.90\textwidth]{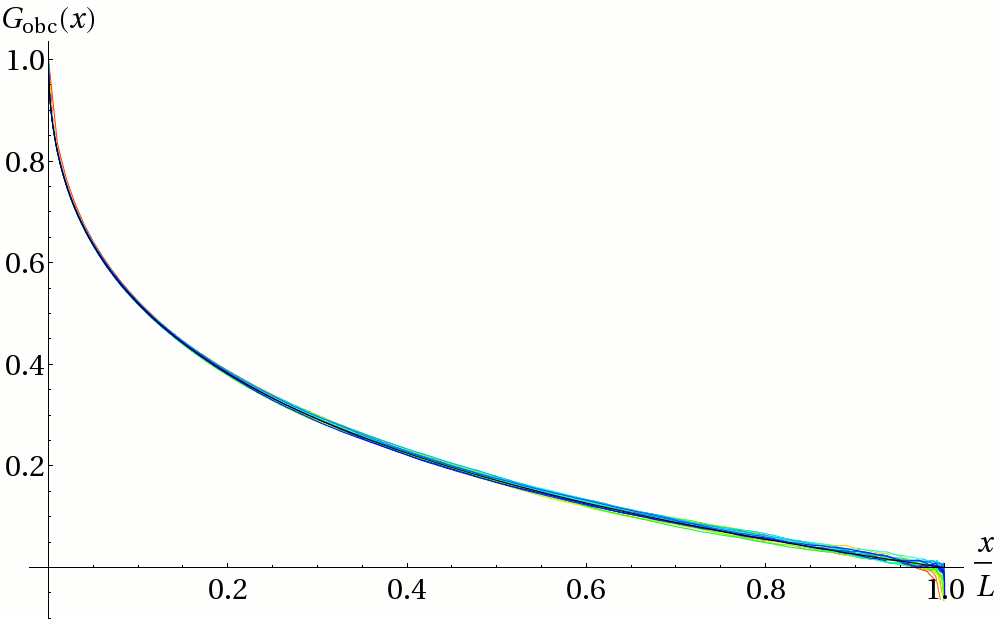}
\end{center}
\caption[]{The correlation function at the critical point with open
  boundary conditions. Top: increasing sizes are represented from red
  ($L=100$) to blue ($L=6500$); Bottom: the same experimental curves,
  rescaled to show the agreement with the theoretical function,
  equation (\ref{eq.hatGteo}), regardless of the size.}
\label{fig:dim1pcOpen}
\end{figure}

In Fig.~\ref{fig:dim1pcOpen} we show the correlation functions at criticality for various sizes. 
In this case each numerical point has been obtained by using $10^4$ instances for the positions of the blue points. All curves are trivially mapped one onto the others by the simple rescaling $r\to r/L$.
That is
\be
G(r; L, 0) = G\left( \frac {r}{L} \right) \, .
\ee
This means that in~\reff{scaling}  we can set $\alpha = 0$. 

In order to estimate the exponent $\nu$, notice that, if in the scaling region, that is $|t| < \overline{t}(L)$, the relation~\reff{scaling} holds, then
\begin{align}
I(r, L) \, := \, L^{-\alpha} \int_0^{\overline{t}(L)} G(r; L, t)\, dt & \, = \,   L^{-\frac{1}{\nu}} \int_0^{\overline{t}(L)\, L^\frac{1}{\nu}} F \left( \frac {r}{L}, z  \right)  dz \label{integral}\\
& \, \approx \,  L^{-\frac{1}{\nu}} \int_0^\infty F \left( \frac {r}{L}, z  \right) dz
\end{align}
because $F$ vanishes rapidly with increasing $z$.
As at fixed $r/L$ the integral of $F$ simply provides a constant, this
expression shows a dependence on $L$ which determines $\nu$.

\begin{figure}[t!]
 \begin{center}
\includegraphics[width=0.90\textwidth]{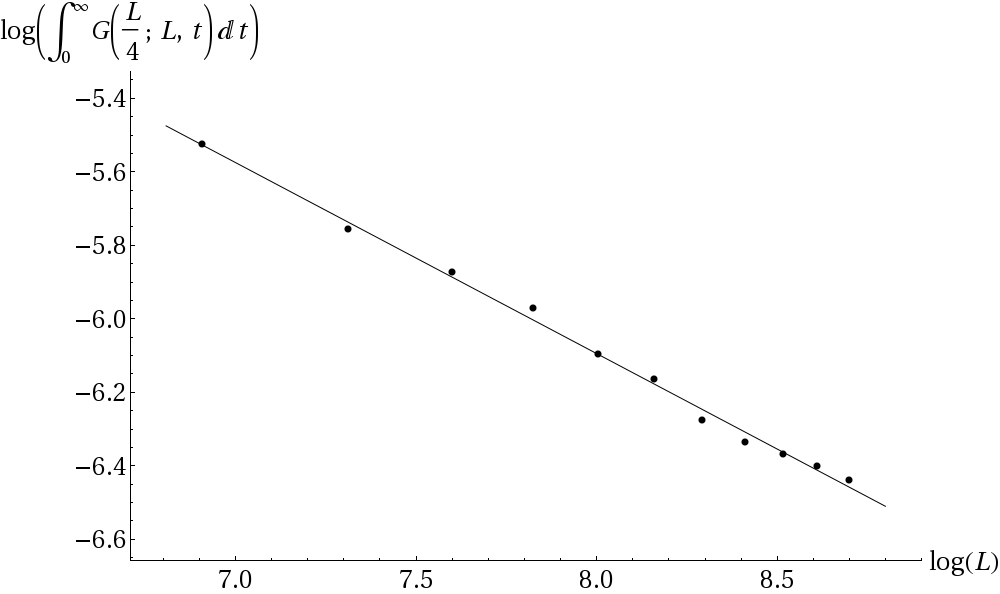}
\end{center}
\caption[]{$\log I(L/4, L)$, defined by~\reff{integral}, for different values of $L$. }
\label{fig:logI}
\end{figure}

In Fig.~\ref{fig:logI} we report the evaluation of $\log I(r, L)$,
defined by~\reff{integral}, for different values of $L$ at the point
in which $r = L/4$. The integral has been evaluated through a
polynomial interpolation in $t$ among the numerical values we had
determined. We find
\be
\nu = 1.95 \pm 0.05 
\ee

\begin{figure}[t!]
 \begin{center}
\includegraphics[width=0.90\textwidth]{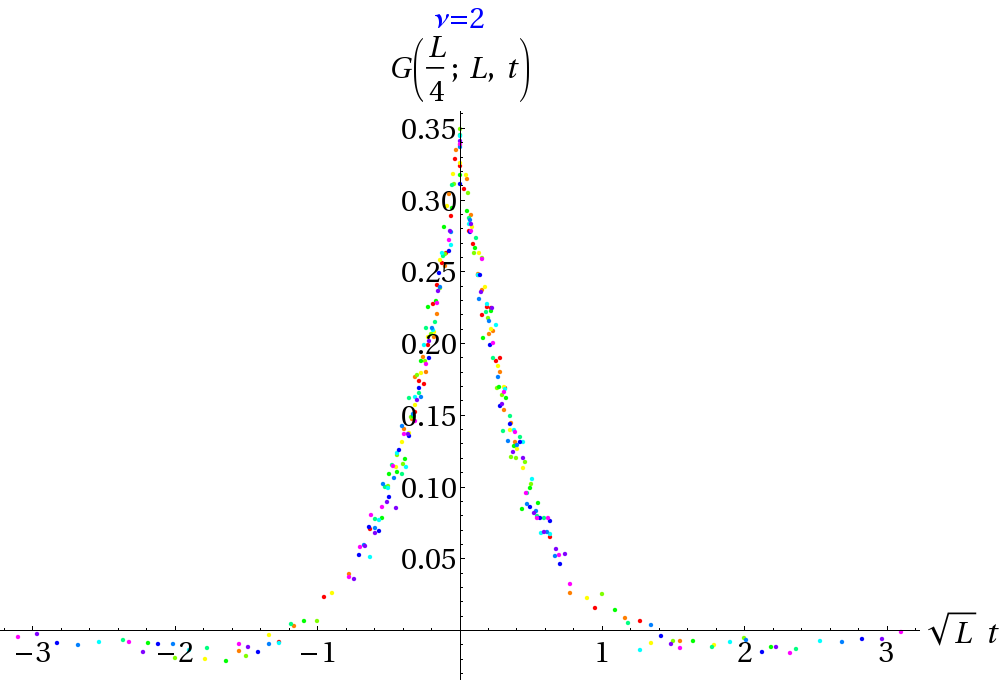}
\end{center}
\caption[]{The correlation function from different values of $L$ at $t$ at $r/L=1/4$ as a function of $\sqrt{L}\, t$.}
\label{fig:Gnu2}
\end{figure}

In Fig.~\ref{fig:Gnu2} we plot the correlation function at $r=L/4$ as a function of $\sqrt{L}\, t $. All the points obtained from different values of $L$, large enough, and $t$, in the scaling region, fall, approximately, on the same curve.

We have performed a similar analysis for the case $\espo<1$, where, in
contrast to the case $\espo>1$ there is no reason to expect that the
values of the critical indices do not depend on $\espo$. Indeed, while
we find always the same exponent $\nu$, the exponent  $\alpha$ shows
the differences  summarised in Table~\ref{tabella}.
\begin{table}[t!]
\begin{center}
\begin{tabular}{|l|c|}
\hline
$\espo$ & $-\alpha$  \\
\hline\hline
1 & 0.00 $\pm$ 0.03  \\
0.95 & 0.08 $\pm$ 0.04  \\
0.90 & 0.18 $\pm$ 0.04  \\
0.85 & 0.28 $\pm$ 0.04  \\
0.80 & 0.37 $\pm$ 0.04  \\
0.75 & 0.47 $\pm$ 0.04  \\
\hline
\end{tabular}
\end{center}
\caption{Numerical estimates of the critical exponent $\alpha$ in the region $\espo <1$. }
\label{tabella}
\end{table}
The exponent $\alpha$ has been computed by using the value of the
two-point correlation function at distance $r = L/4$ and the scaling
ansatz~\reff{scaling} at criticality.
We find approximately
\be
\alpha \approx  - 2 ( 1 -\espo )
\ee
in the region $\espo \in [0.75, 1]$.

\subsection{Analytical predictions at criticality}

Recall that, from the fact that the solution is ordered, we just have,
for $i=1,\dots, L$
\be
\varphi_i = b_i - r_i = b_i - i + \frac{1}{2} \, .
\ee
The collection of $\varphi_i$'s, for $i$ labeled in order, can be
transformed into a stochastic function from $[0,1]$ to $\mathbb{R}$,
by setting $\varphi(s) = \varphi_i$ for $s \in [(i-1)/L,i/L]$.
The parameter $s\in [0,1]$ is a sort of {\em time} variable, for the
evolution of a random walk.

Given that there are $L$ blue points in the interval $[0, L]$, the
probability to find the $i$-th blue point in the interval $[y, y+dy]$
is given by
\be
P_i(dy)  = \frac{\frac{y^{i-1}}{(i-1)!} \frac{(L-y)^{L-i}}{(L-i)!}}
{\frac{y^L}{L!}  }\,dy  = \binom{L}{i}\, \left( \frac{y}{L} \right)^i
\left( 1- \frac{y}{L} \right)^{L-i} \frac{i}{y} \,dy =
B_i \left(L; \frac{y}{L} \right)\, \frac{i}{y} \, dy
\ee
where $B_i(n; p)$ is the binomial distribution, for getting $i$ `head'
when tossing $n$ times a biased coin with probability $p$ for `head'.
In the limit of large $L$, by keeping fixed the ratio $s=y/L$, we
get by the central limit theorem that
\be
B_i \left(L; s \right) \to \frac{ e^ { - \frac{(i-y)^2}{2 \,
      L s \left( 1 - s \right) }}}{\sqrt{ 2 \pi \,
    L s \left( 1 - s \right) } }
\ee
which tells us that the difference $i - y$ is of order $\sqrt{L}$ so
that by the change of variables
\be
\varphi_i  = \sqrt{L} \, x_i + \frac{1}{2}
\ee
we get the probability distribution 
\be
p_{B(s)} (x) \, = \, \frac{ e^ { - \frac{x^2}{2 \,s \,( 1 - s)
}}}{\sqrt{ 2 \pi \,s \,( 1 - s) } }
\ee
which is the Gaussian probability distribution of a Brownian bridge
$B(s)$ over the interval $[0, 1]$ (see Fig.~\ref{fig:brownianBridge}).
Precise definitions and further details can be found in
Appendix~\ref{app}.

Essentially the same calculation can be performed for the joint
probability distribution for displacements from different blue points
to see that it always provides, in the limit of large $L$,  the joint
distribution at different times of a Brownian bridge. 
\begin{figure}[t!]
\begin{center}
\includegraphics[width=0.9\textwidth]{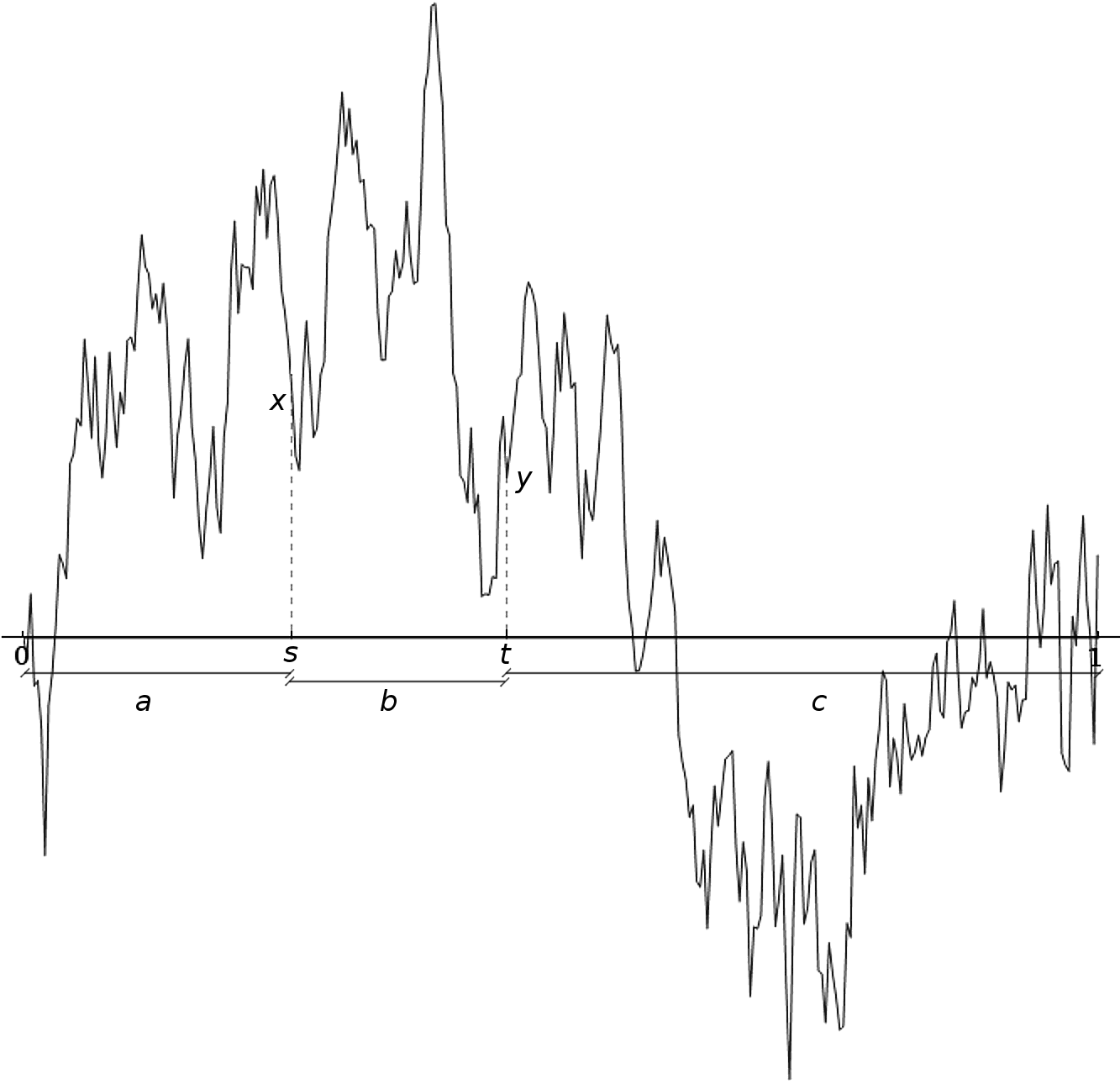}
\end{center}
\caption{Schematic representation of a Brownian bridge $B$ over the
  interval $[0, 1]$, and its value at the intermediate times $s$ and~$t$.}
\label{fig:brownianBridge}
\end{figure}

Remark that this result would remain unchanged for the Poisson-Poisson
matching. In that case, also the red points would be distributed like
the blue points, and the difference of two random variables
distributed according to the same binomial distribution $B(L; y/L)$ is
still a binomial with distribution $B(2 L; y/L)$. In order to converge
to the same continuum limit we must rescale our variables as
\be
\varphi_i  = \sqrt{2\,L} \, x_i 
\ef,
\ee
which explains the discrepancy, in certain factors $2$, of our results
w.r.t.\ analogous calculations, present in the literature, for the
Poisson-Poisson case.

As a first consequence of the distribution function that we have
obtained, we can evaluate the cost of the optimal Grid-Poisson
matching:
\begin{align}
L^{-\frac{\espo}{2}}\, E_{\rm opt} \,\to \,  \int_0^1 ds\, {\mathbb E}
(x^\espo(s)) \, = \,  & \int_0^1 ds\, \frac { [ 2 s
    (1-s)]^\frac{\espo}{2} }{\sqrt{\pi}} \, \Gamma\left( \frac{ 1 +
  \espo}{2} \right) \\
\, = \,  & \sqrt{ \frac{2^\espo}{\pi} }\,  \frac{ \Gamma^2 \left(
  \frac{\espo}{2} + 1 \right) } {\Gamma(\espo+2)} \,  \Gamma\left(
\frac{ 1 + \espo}{2} \right) \\
\, = \,  &  \sqrt{ \frac{2^{-\espo}}{\pi} }\, \frac{\Gamma\left(
  \frac{\espo}{2} + 1  \right) }{p+1}
\end{align} 
for $\espo >1$, so that, in particular, for $\espo=2$ we get
\be
L^{-1}\, E_{\rm opt} \,\to \,  \int_0^1 ds\, s \, (1-s) = \frac{1}{6} \, .
\ee

Now, let us consider two intermediate times, $s$ and $t$, with $0 < s
< t < 1$ (see again Fig.\ \ref{fig:brownianBridge}). The probability
that the process started at the origin arrives at $x_1$ after a time
$s$ is that of a Wiener process, so that it is a Gaussian with zero
mean and variance $s$:
\begin{equation}
p_{W(s)}(x_1) = \frac{1}{\sqrt{2{\pi}s}}\,  e^{-\frac{x_1^2}{2s}}.
\end{equation}
Similarly, to move from $x_1$ to $x_2$ in the interval $(t-s)$:
\begin{equation}
p_{W(t-s)}(x_2-x_1) = \frac{1}{\sqrt{2{\pi}(t-s)}}\, e^{-\frac{(x_2-x_1)^2}{2(t-s)}}
\end{equation}
and, finally, to move from $x_2$ to $0$ in the interval $(1-t)$:
\begin{equation}
p_{W(1-t)}(x_2) = \frac{1}{\sqrt{2{\pi}(1-t)}}\, e^{-\frac{x_2^2}{2(1-t)}}.
\end{equation}
Since the distribution is Gaussian, which means that the joint
distribution has the form~\reff{eq:gaussA}
discussed in the Appendix
\be
p_A(x_1, x_2) = \sqrt{\det A}\ \frac{e^{-\frac{1}{2}\sum_{i=1}^{2}x_i A_{ij} x_j}}{2\pi},
\ee
by a change of parameters, if we consider the three segments of length 
\begin{equation}
\renewcommand{\arraystretch}{1}\addtolength{\tabcolsep}{1pt}
\begin{array}{ll}
 a &=s  \\
 b &=t-s  \\
 c &=1-t ,
\end{array}
\label{eq:par}
\end{equation}
we get that the matrix $A$ is given by
\begin{equation}
  A = \left( \begin{array}{cc}
            \frac{1}{a} + \frac{1}{b} & -\frac{1}{b}\\
            -\frac{1}{b} & \frac{1}{c} + \frac{1}{b}
\end{array}
 \right)
 \label{eq:matrAnp}.
\end{equation}
with
\begin{equation}
 \det A = \frac{a + b + c}{a b c}.
\end{equation}


\noindent
Let us now investigate the correlation function.
Given the continuous variable
\be
\sigma(s) := \frac{\varphi(s)}{|\varphi(s)|} = \sgn (\varphi(s))
\ee
for $s\in [0,1]$, 
we shall look at the correlation function 
\begin{equation}
G(s, t) = 
\left\langle
\sigma(s) \sigma(t) 
\right\rangle
=
\left\langle
\sgn(\varphi(s)  \varphi(t))
\right\rangle
\ef.
\end{equation}
Let us assume $s<t$, rename $s=a$ and $t=a+b$, and use $c$ as a synonim
of $1-a-b$. We can study the slightly more general quantity, function
of $a$, $b$ and $c$ with no constraint $a+b+c=1$
\begin{align}
G(a, b, c) &= \int \! \int  \mathrm{d} x \: \mathrm{d} y \, p_{A}(x,y)
\sgn(x\cdot y) \nonumber \\
&= \int \! \int  \mathrm{d} x \: \mathrm{d} y \, \sqrt{2 \pi}
\ \sqrt{a + b + c} \  \frac{e^{-\frac{x^2}{2a}-\frac{(x-y)^2}{2b}
    -\frac{y^2}{2c}}}{\sqrt{2 \pi a} \ \sqrt{2 \pi b} \ \sqrt{2 \pi
    c}} \ \sgn(x\cdot y)
\end{align}
If we define 
\begin{equation}
\alpha(a, b, c) := \int_{x\geq0} \! \int_{y\geq0} \! \mathrm{d} x \: \mathrm{d} y \ p_{A}(x,y)  
\label{eq:alfa}
\end{equation}
and
\begin{equation}
\beta(a, b, c) := \int_{x\geq0} \! \int_{y\leq0} \! \mathrm{d} x \: \mathrm{d} y \  p_{A}(x,y),  
\end{equation}
then
\begin{equation}
G(a, b, c) = 2 \,\alpha(a, b, c) - 2\, \beta(a, b, c).
\end{equation}
In addition, since $p_{A}(x,y)$ is a normalised Gaussian, we know that
\begin{equation}
2\, \alpha(a, b, c) + 2\, \beta(a, b, c) = 1,
\end{equation}
then
\begin{equation}
G(a, b, c) = 4\, \alpha(a, b, c) - 1.
\end{equation}
By performing the integral \eqref{eq:alfa}, we find
\begin{equation}
 \alpha(a, b, c) = \frac{1}{4} + \frac{1}{2 \pi} \arctan \sqrt{\frac{a c}{b(a+b+c)}},
\label{eq:risalfa1}
\end{equation}
and then
\begin{equation}
G(a, b, c) = \frac{2}{\pi} \arctan \sqrt{\frac{a c}{b(a+b+c)}}
\ef.
\label{eq:g2np}
\end{equation}
Specialising to
$a+b+c=1$, this simplifies to
\begin{equation}
G(a, 1-a-c, c) = \frac{2}{\pi} \arctan \sqrt{\frac{a c}{1-a-c}}
\ef.
\label{eq:g2npSimpl}
\end{equation}
If we keep the distance $b=t-s$ between the two points constant, and we
calculate the mean over the interval $[0, 1]$, we finally obtain
\begin{equation}
\label{eq.hatGteo}
G_{\rm obc}(b) = 
\frac{1}{1-b}
\int_0^{1-b}
\mathrm{d} a
\;
G(a, b,1-a-b)
= \frac{1 -
  \sqrt{b}}{1 + \sqrt{b}}.
\end{equation}
We found an excellent  agreement of the theoretical predictions with
the numerical data, even at sizes as small as $L=100$. 
This can be seen in Fig.~\ref{fig:dim1pcOpen}.

\section{Periodic boundary conditions}

In this section we analyse the one-dimensional Matching Problem with
periodic boundary conditions. All along the section, the indices are
considered modulo $L$ (e.g., $r_i$ and $r_{i+L}$ are the same
red-point coordinate).

\subsection{Properties of the optimal matching}

Also in the realisation of the problem with periodic boundary
conditions, criticality is obtained when the two sets of points have
the same cardinality.  However, we no
longer have here the trivial characterisation of the optimal solution for $\espo>1$. We
have a result analogous to Proposition \ref{prop.sec2}, that is
however more subtle and complicated.  Again, we compare
subconfigurations of candidate solutions, but, differently from the
open-boundary case, we need to consider \emph{triples} of points,
instead of pairs.

For an ordered triple of points $(x,y,z)$ on an (oriented) circle, we
say that it is \emph{cyclically oriented} if $x$, $y$ and $z$ appear
on the circle in (say) counter-clockwise order. Clearly, any other
permutation of the three points will be cyclically oriented, or not,
depending on the signature of the permutation. For two triples, we
say that they are \emph{cyclically co-oriented} if they are oriented
in the same direction (clockwise or counter-clockwise).

A maximal matching $\pi$ is said to be \emph{cyclic} if, for all
triples of distinct edges $(i_1,j_1)$, $(i_2,j_2)$, $(i_3,j_3)$ in
$\pi$, the two triples $(i_1,i_2,i_3)$ and $(j_1,j_2,j_3)$ are
cyclically co-oriented.

The following lemma is what we need to obtain a characterization of
the solution for the matching problem when $p\ge 1$.

\begin{prop}
Consider the matching problem for distinct points on the circle of
unit length ${\cal S}^1$ with distance between two points given by the
minimal length along the two possible connecting paths.  If $\espo>1$,
the optimal matching $\pi_{\rm opt}$ is unique and cyclic. If
$\espo=1$, there exists a cyclic optimal matching.
\end{prop}

\prf 
We proceed by absurd, assuming that an optimal matching $\pi$ has a
triple of edges which are not cyclically co-oriented, and showing that
a transposition of two of these edges decreases the cost if $p>1$. The
statement for $p=1$ then follows by continuity.

Call $(r_1, r_2, r_3)$ the positions of the three red points, in
cyclic order, and $(b_1, b_2, b_3)$ those of the blue points, also in
cyclic order, here given as reals in $[0,1)$ and intended modulo
1. The antipodal positions $(\bar{r}_1,\bar{r}_2,\bar{r}_3)$ and
$(\bar{b}_1,\bar{b}_2,\bar{b}_3)$ are the same lists, shifted by
$\frac{1}{2}$, and taken modulo 1. This makes $12$ points on the circle, say
for simplicity all distinct, and 12 intervals (the case of points at
antipodes corresponds to the limit of some interval having zero
length, that just simplifies the treatment).

Such a structure is equivalently encoded by an ordered $12$-tuple
$(x_1,\ldots,x_{12})$ of elements in $[0,1)$, such that
$x_{i+6}=x_i+\frac{1}{2}$, and by a string of $12$ elements
in $\{\textrm{red},\textrm{blue},\textrm{white}\}$ such that red and blue
have three preimages (here white stands for antipodal, regardless from
the colour of the antipode). We call such a string a \emph{pattern},
and denote by $(a_1,a_2,\ldots,a_6)$ the lengths of the intervals,
i.e.~$(x_2-x_1,x_3-x_2,\ldots, \frac{1}{2}+x_1-x_6)$.  These
parameters are subject to $a_1+\cdots+a_6=\frac{1}{2}$, but, given the
homogeneity of the cost function, we can safely ignore this
constraint.

We can assume, without loss of generality, that the first of the 12
points is a red point, i.e.\ restrict to patterns starting with
``red''. This makes $320=\binom{5}{2} 2^5$ possible patterns.
Using the $D_3$ dihedral symmetry of the problem at hand,
this can be reduced to $72$\,\footnote{That is slightly more than
  $320/6$, due to the fact that we save less by symmetry for
  configurations having a non-trivial group of automorphisms.}.

The reason for considering the antipodal points is the fact that the
distance is given by an ``if'' condition, on the lengths of the two
paths. Once the antipods are taken into account, we see that the two
paths have length of the form $a_l+ \cdots+a_{l+s}$ and $a_l + a_{l+s}
+ 2 a_{l+s+1}+ \cdots + 2 a_{l+6}$ (all the indices are modulo 6), for
$1\leq s \leq 5$ and $1\leq l \leq 6$, and the ``if'' statement
trivialises. So, for a given pattern, all the 9 relevant distances
from $r_i$ to $b_j$ are easily determined, simultaneously for all
choice of non-negative parameters $\{a_l\}$.

The comparison of distinct permutations is made a bit simpler by the
fact that two permutations with opposite signature in $\mathfrak{S}_3$
always differ by a simple transposition, so that one expression is in
common.

What we manage to prove is something slightly stronger than the claim
in the proposition. Namely, not only we prove that, for all patterns
$P$, non-cyclic permutations $\{(132),(213),(321)\}$, and choice of
parameters $\{a_l\}$, there exists a cyclic permutation in
$\{(123),(231),(312)\}$ of lower cost, but also that the choice of
such cyclic permutation can be made uniform for a given pattern,
regardless of the non-negative values of the $\{a_l\}$'s. Furthermore,
when comparing a cyclic and a non-cyclic permutation, the bound occurs
through two mechanisms only.

\begin{itemize}
\item One or more variable $a_l$ does not appear in the two non-common
  expressions for the distances in the cyclic permutation, while it
  appears in the expressions for the distances in the non-cyclic
  one. If these variables are set to zero, the three expressions for
  the distances become identical in the two permutations.
\item Possibly after setting some variables to zero as in the previous
  case, the two non-common expressions take the form $\{A+B,B+C\}$,
  for the cyclic permutation, and $\{A+B+C,B\}$, for the non-cyclic
  one. Then, the cyclic permutation has a lower cost because
  $(1+x)^p+(1+y)^p<1+(1+x+y)^p$ for all $p>1$ and $x$, $y>0$.
\end{itemize}
The second mechanism is compatible with the emergence of degeneration
of the optimal solution at $p=1$, as in this case, of course,
$(1+x)+(1+y)=1+(1+x+y)$.

We successfully checked the $72$ patterns, and three non-cyclic
permutations per pattern, by computer, under the ansatz that the bound
was of one of the two forms above.

Just to give an example, consider the pattern 
$[\bulB \bulX \bulX \bulX \bulB \bulW \bulX \bulB \bulW \bulW \bulX \bulX]$
(where we used 
$\bulB$, $\bulW$ and $\bulX$ for red, blue and white, respectively),
and just call $a,b,\ldots,f$ the parameters $a_l$.
We have
\[
\begin{array}{l|l}
\multicolumn{2}{c}{\textrm{cyclic}}
\\
\hline
    & a+b+c+d+e \\
123 & e+f+a+b \\
    & b+c \\
\hline
    & c+d+e+f \\
231 & e+f+a+b+c \\
    & f+a \\
\hline
    & d+e+f \\
312 & e \\
    & b \\
\end{array}
\qquad
\begin{array}{l|l}
\multicolumn{2}{c}{\textrm{non-cyclic}}
\\
\hline
    & a+b+c+d+e \\
132 & e+f+a+b+c \\
    & b \\
\hline
    & c+d+e+f \\
213 & e \\
    & b+c \\
\hline
    & d+e+f \\
321 & e+f+a+b \\
    & f+a \\
\end{array}
\]
The non-cyclic permutation $(213)$ is bounded by $(312)$, as $c$ does
not appear in the expressions for the latter, and, after setting $c\to
0$, the two unordered lists of expressions do coincide, thus the first
criterium applies.

The non-cyclic permutation $(132)$ is bounded by $(123)$, as the first
expressions coincide, and the two remaining ones have the form
$(A+B+C,B,A+B,B+C)$ with $B=b$, and $\{A,C\}=\{e+f+a,c\}$, thus the
second criterium applies.

The non-cyclic permutation $(321)$ is bounded by $(312)$, as the first
expressions coincide, and, for the two remaining ones, after setting
$f=a=0$, we have the form $(A+B+C,B,A+B,B+C)$ with $B=0$, and
$\{A,C\}=\{b,e\}$, thus the second criterium applies.

As mentioned above, all the other patterns are treated similarly.
\qed

\begin{prop}
When $|\mathcal{R}|=|\mathcal{B}|=L$, the cyclic maximal matchings are
all and only the permutations of the form
$(i,i+1,\ldots,L,1,2,\ldots,i-1)$.
\end{prop}

\prf Let $i \neq j$. Either $\pi(i) - \pi(j) = i - j$ modulo $L$ for
all pairs, or there exists at least one pair of indices $i$ and $j$
such that  $\pi(i) - \pi(j) \neq i - j$ modulo $L$.
In the second case, consider the interval $I= (i, i+1, \cdots , j)$
and $J = (\pi(i), \pi(i)+1, \cdots, \pi(j))$. As these intervals have
different cardinality, there must exists a $k$ such that either $k\in
I$ and $\pi(k) \not\in J$ or $k\not\in I$ and $\pi(k)\in J$. Therefore
the triples $(i, j, k)$ and $(\pi(i), \pi(j), \pi(k))$ are not
cyclically co-oriented.
\qed

This means that, if we label the points in the sets $\cal R$ and $\cal
B$ in increasing counter-clockwise order, in the unique optimal solution
(for $p>1$, or one optimal solution, for $p=1$) our function $\phi_i$
has the form
\be
\varphi_i = b_i - r_{i-\ell}
\label{eq.65346545}
\ee
with a particular $\ell \in \{0, \dots, L-1\}$.

\subsection{Numerical results}

\begin{figure}[t!]
 \begin{center}
\includegraphics[width=0.9\textwidth]{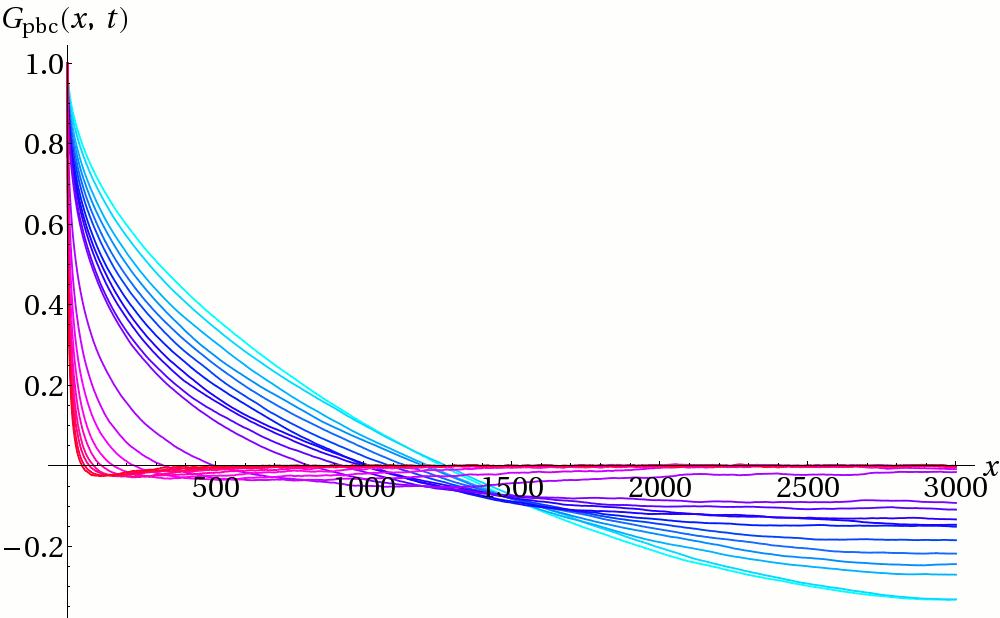}
\end{center}
\caption[]{The correlation function near the critical point at size 6000 with periodic boundary conditions for the case $p=2$. 
From cyan to purple, $t= $0, 0.001, 0.002, 0.003, 0.004, 0.005, 0.006, 0.007, 0.008, 0.009, 0.01, 0.02, 0.03, 0.04, 0.05, 0.06, 0.07, 0.08, 0.09, 0.1. }
\label{fig:dim1fuoriPcCerchio}
\end{figure}

In Fig.~\ref{fig:dim1fuoriPcCerchio} we report the correlation
function $G(r; L, t)$ for various choices of the parameter $t$ when
the size of the system is $L=6000$.
Each curve is the mean over $10^3$ instances.
The function is even under the parity $r \to L- r$ so that we plot it
only in the interval $[0,L/2]$.
The function $G(r; L, t)$ still presents two ranges of behaviour. If
$|t|<\bar{t}(L)$, with $\bar{t}(6000)\approx 0.01$, the function is
decreasing with $r$, and has, therefore,  a minimum at
$r=\frac{L}{2}$. 
However, contrarily to the case of open boundary conditions, 
this is no longer positive definite.
When $|t|>\bar{t}(L)$, the shape is different. It reaches a minimum at
an intermediate value $r'$, then goes up again approaching zero as
$r\to \frac{L}{2}$.

\begin{figure}[t!]
 \begin{center}
\includegraphics[width=0.9\textwidth]{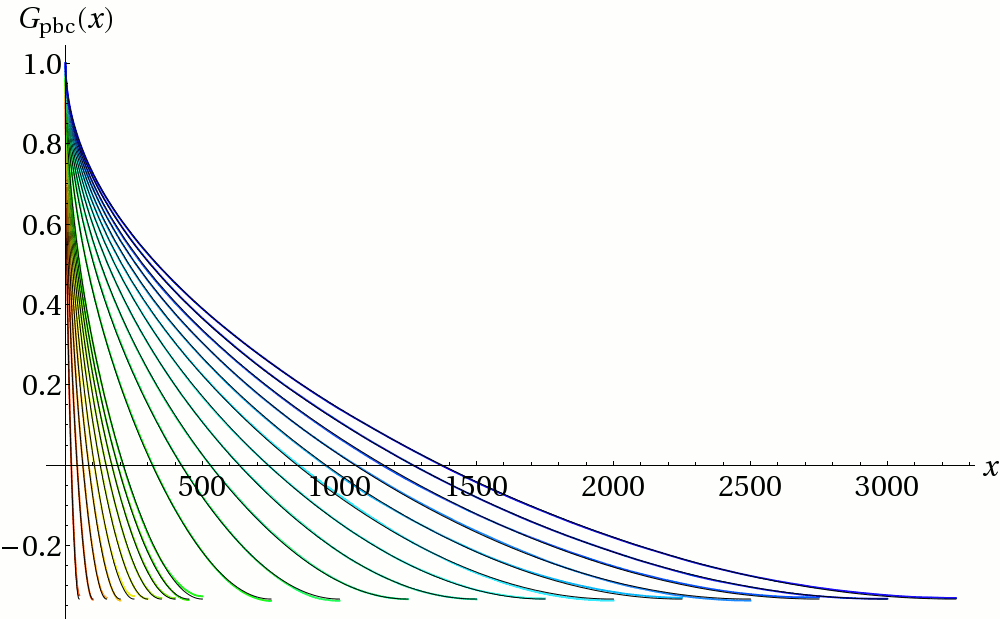}\\[0.7 cm]
\includegraphics[width=0.9\textwidth]{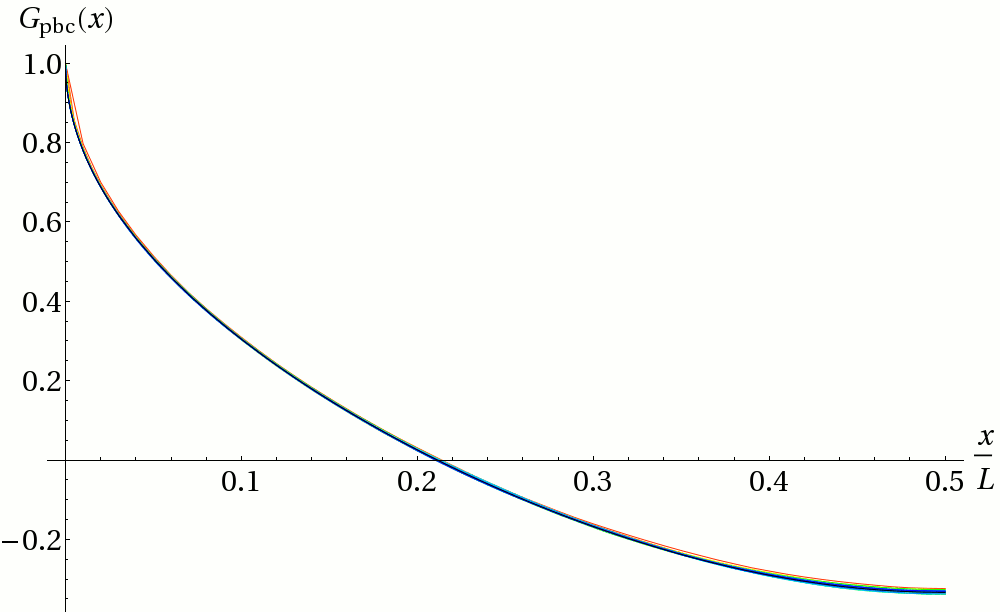}
\end{center}
\caption[]{The correlation function with periodic boundary conditions at the critical point  for the case $p=2$. Top: increasing sizes are represented from red ($L=100$) to blue ($L=6500$); Bottom: the same experimental curves, rescaled to show the agreement with the theoretical function, regardless of the size.}
\label{fig:dim1pcPer}
\end{figure}

In Fig.~\ref{fig:dim1pcPer} we show the correlation functions at
criticality for various sizes. Each numerical point has been obtained
by using $10^4$ instances for the positions of the blue points. Again,
all curves collapse under the simple rescaling $r\to r/L$.

\begin{figure}[t!]
 \begin{center}
\includegraphics[width=0.90\textwidth]{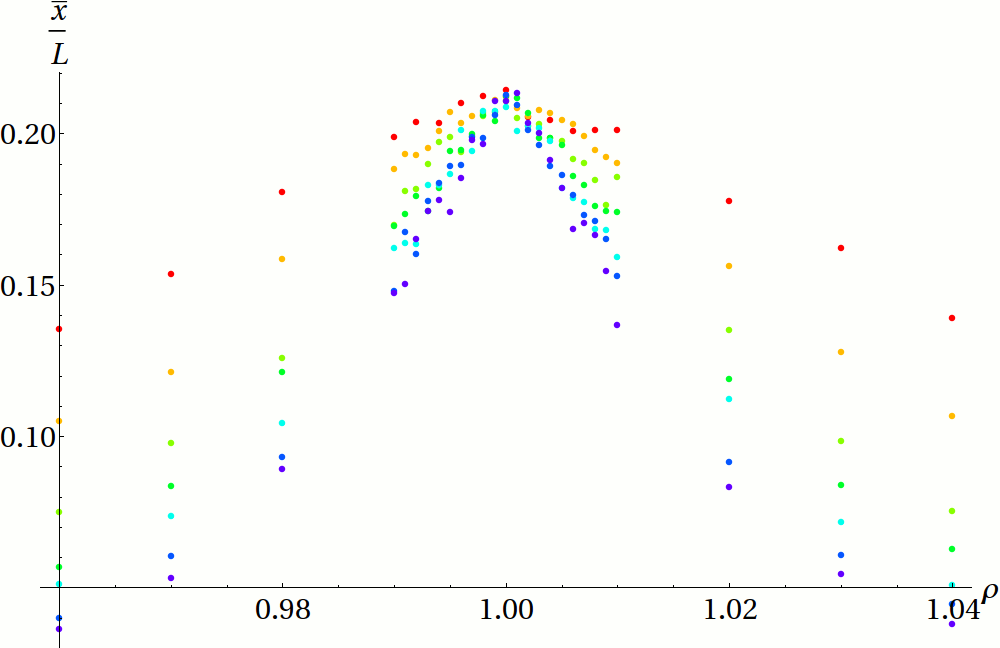}
\end{center}
\caption[]{The experimental rescaled intersection with the $x$-axis (\(\bar{x}_L/L\)) as a function of the {\em reduced temperature} $t$. Colours from red to purple for sizes $L=500, 1000, 2000, 3000, 4000, 5000,6000$.}
\label{fig:xiDim1}
\end{figure}

When $|t|<\bar{t}(L)$, if we define  \(\bar{x}_L \) as the point where
the curve for the size $L$ has a zero, we find that \(\bar{x}_L(t)\)
has a maximum at criticality \(t=0\) (see Fig.~\ref{fig:xiDim1}).
The value of $\bar{x}_L/L$ at criticality as a function of the size
$L$ is almost constant, we get 
\begin{equation}
\frac{\bar{x}_L(t=0)}{L} \approx 0.2117 \pm 0.0004 \, .
\label{eq:xbar1dim}
\end{equation}
%
%
%

\begin{figure}[t]
 \begin{center}
\includegraphics[width=0.9\textwidth]{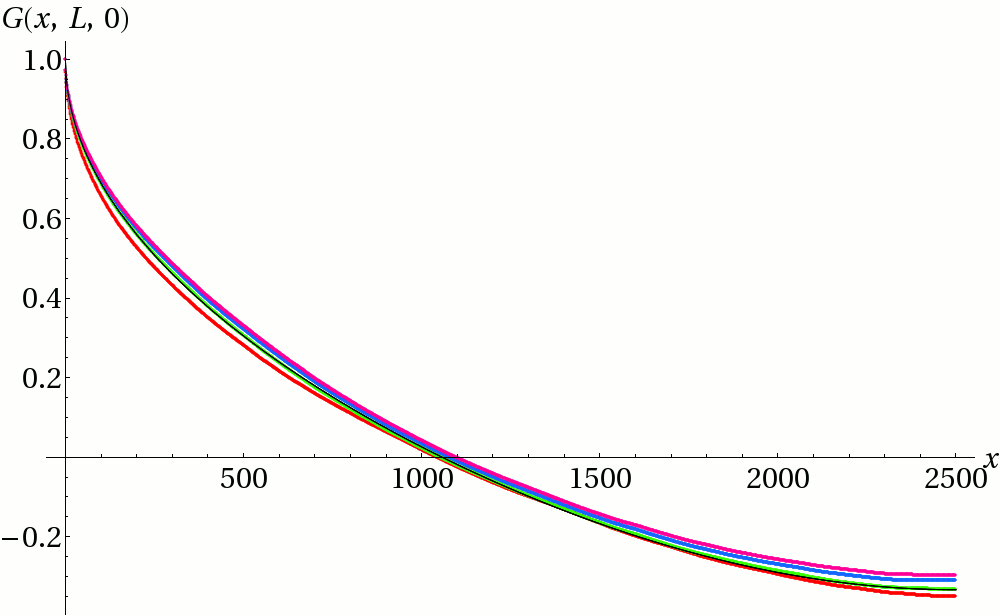}
\end{center}
\caption[]{Comparison of the correlation function at criticality, with
  different values of $\espo$ ($L=5000$). Lines in red, green, blue
  and purple correspond to $\espo= 1, 2, 3, 4$. The black thinner
  curve is the plot of the analytic function predicted for $\espo=2$.}
\label{fig:cfrGamma}
\end{figure}

We have investigated numerically what happens when we change the
exponent $\espo$ which appears in the cost function, by looking also
at the values $\espo= 1, 3, 4$.  The shift which determines the
optimal solution is in general different from what we have at
$\espo=2$, and this difference has a consequence on the correlation
function.  We observe however that the relative variation of the curve
is quite small.  Fig.~\ref{fig:cfrGamma} presents these results. We
plot the  correlation function at criticality, that is $t=0$, for the
size $L=5000$. Each curve is the mean over $10^3$ instances. 

\subsection{Analytical predictions at criticality}

Recall that the optimal configuration has the structure described in
equation (\ref{eq.65346545}), i.e.\ there exists an integer $\ell$
such that $\pi_{\rm opt}(i)=i-\ell$, and thus $\varphi_i = b_i -
r_{i-\ell}$. We can define $\varphi(s)$ analogously to what was done in
the previous section, and pass to the continuum limit.  Then, after
the rescaling with $L$, the integer shift $\ell$ becomes a real
variable $\lambda$\,\footnote{Precisely, $\lambda \in
[-\frac{\sqrt{L}}{2},\frac{\sqrt{L}}{2}]$, and the interval converges
to $\mathbb{R}$ in the limit.}
so that our process is a
Brownian Bridge, vertically translated by $\lambda$
\be
\varphi_\lambda(s) = B(s) - \lambda
\ee
By definition, the cost is
\be
E[\varphi_\lambda]  = \int_0^1 ds\, |\varphi_\lambda(s)|^\espo  =
\int_0^1 ds\, \left|B(s) - \lambda\right|^\espo
\ef.
\ee
We have a simple necessary condition for
optimality, that corresponds to stability w.r.t.\ the application of
an elementary cyclic rotation $i \to i \pm 1$
\be 
\frac{d}{d \lambda} E[\varphi_\lambda]  = 
-\espo \int_0^1 ds\, \frac{|\varphi_\lambda
  (s)|^{\espo}}{\varphi_\lambda (s)}  = 0 
\ef.
\label{eq.1866453}
\ee
For a fixed configuration of points,
this equation is complicated for generic $\espo$. For $\espo$ even, it
is a real-valued polynomial of degree $\espo-1$, and has in general
$\espo-1$ roots, and, more precisely, $2k-1$ real roots and
$\frac{1}{2}\espo-k$ pairs of complex-conjugate roots. Thus, we have at least
one real root, and the global minimum must be achieved at one of these
roots. In fact, at generic $\espo\geq 1$ the equation has a unique
real root, as can be proven by a simple argument. First of all, call
$\mu_B(x)$ the density induced by the Brownian Bridge,
i.e.\ $\mu_B(x)$ is the derivative of $\int \mathrm{d}s \;
\theta(B(s)-x)$. Then, the energy reads
\be
E[\varphi_\lambda]  = \int d\mu_B(x)\, 
\left|x - \lambda\right|^\espo
\ef,
\ee
while the stability condition reads
\be
\int d\mu_B(x)\, 
\left|x - \lambda\right|^{\espo-1}
\sgn(x-\lambda)
=0
\ef.
\ee
This equation has always an odd number of real roots. For $\espo=1$,
this is obvious. For $\espo>1$ the asymptotics for $\lambda \to
\pm \infty$ is $\sim \pm|\lambda|^{\espo-1}$.
A further derivative gives
\be
-(\espo-1)
\int d\mu_B(x)\, 
\left|x - \lambda\right|^{\espo-2}
\sgn(x-\lambda)^2
=
-(\espo-1)
\int d\mu_B(x)\, 
\left|x - \lambda\right|^{\espo-2}
\ef,
\ee
which has definite sign, thus proving the convexity in $\lambda$ of 
$E[\varphi_\lambda]$. So, the condition (\ref{eq.1866453}) is
necessary and sufficient for optimality.





When $\espo=2$ this relation is just linear. In this case the
optimality condition is trivially solved by
\be
\lambda = \int_0^1 ds\, B(s) \,  
\ee
It is useful to define the area under the path after time $t$
\be
B^{(-1)}(t)  := \, \int_0^t ds\, B(s) 
\ee
so that the previous equation just reads
$\lambda  = B^{(-1)}(1)$.

We therefore deduce that our solution converges in the continuum to
the process
\begin{equation}
\varphi(s) = \,  B(s) - B^{(-1)}(1)  \, 
\label{eq:varpbc}
\end{equation}
which is a linear combination of the elementary process $B(s)$. 
The value of $\phi$ at a given coordinate $s$ is, again, a Gaussian
random variable with zero expectation
value. The covariance at two coordinates $(s,t)$ can be easily
calculated as follows:
\begin{equation}
 \cov[B(s) - B^{(-1)}(1) , B(t) - B^{(-1)}(1)] = \frac{1}{12} -
 \frac{1}{2}t(1-t) - \frac{1}{2}s(1-s) + \min(s, t) - s t.
\label{eq:covbb_pbc}
\end{equation}
If we assume $s \leq t$, this expression becomes
\begin{equation}
  \cov[ \varphi(s), \varphi(t)] = \frac{1}{12} - \frac{1}{2}(t-s)(1-(t-s)),
\end{equation}
which, as expected, is translational invariant (i.e., depends only on
$t-s$) and symmetric w.r.t.\ reflection (i.e.,
depends symmetrically on $t-s$ and
$1-(t-s)$).
The expression for the covariance satisfies 
\be
\label{eq.453634765}
\frac{d^2}{ds^2} \cov[ \varphi(s), \varphi(t)]  = -\delta(s-t) +1
\ee
where the ``$+1$'' correction to the customary $\delta$-function is
induced by the fact that it shall balance the latter, as, in presence
of periodic boundary conditions, the integral of the left-hand side of
(\ref{eq.453634765}) on the whole interval must vanish.

As a consequence, in the case of periodic boundary conditions, the
optimal cost for unit length is
\be
L^{-1} E_{\rm opt} = \frac{1}{12} 
\ee
in agreement with the analysis performed in~\cite{clps} for the
Poisson-Poisson matching, which must differ from this result by a
factor two (because of the double contribution to fluctuations, from
red and blue points).

If we define
\begin{equation}
 \tau = t - s
\end{equation}
and 
\begin{equation}
 \eta = \tau(1-\tau) = (t-s)(1-(t-s)),
\end{equation}
the covariance matrix can be written as
\begin{equation}
 C = \left( \begin{array}{cc}
            \frac{1}{12} & \frac{1}{12} - \frac{1}{2}\eta \\
            \frac{1}{12} - \frac{1}{2}\eta & \frac{1}{12}
\end{array}
 \right)
\label{eq:matrCp}
\end{equation}
and therefore in this case the joint probability distribution is still
of the form~\reff{eq:gaussA} discussed in the Appendix, now with the matrix $A$
\begin{equation}
  A = C^{-1} = \frac{1}{\eta(1- 3\eta)} \ \left( \begin{array}{cc}
           1 & -1 + 6\eta\\
            -1 + 6\eta & 1
\end{array}
 \right).
\label{eq:matrAp}
\end{equation}
By comparing \eqref{eq:matrAnp} with \eqref{eq:matrAp}, we find
\begin{equation}
\begin{array}{l}
\!\! b= \frac{\eta(1- 3\eta)}{1 - 6\eta} \\
\!\! a=c=\frac{1- 3\eta}{6}.
\end{array}
\end{equation}
The important difference with respect to the non-periodic case is that
\begin{equation}
 a+b+c = \frac{(1- 3\eta)^2}{3(1 - 6\eta)},
\end{equation}
which is in general  $\neq 1$. Moreover, $b$ and $a+b+c$ can now have a negative sign:
 \begin{equation}
\renewcommand{\arraystretch}{1}\addtolength{\tabcolsep}{1pt}
\begin{array}{l}
   b < 0 \\
a + b + c < 0 
  \end{array}
\ \ \ \mathrm{if} \ \ \ \eta > \frac{1}{6}.
 \end{equation}
If $\eta < \frac{1}{6}$, the result in \eqref{eq:risalfa1} is still valid, while if $\eta > \frac{1}{6}$, we obtain\footnote{Here we have used the trigonometric identities: $\arctan x + \arctan(\frac{1}{x}) = \frac{\pi}{2}$ and $\arctan(-x)=-\arctan x$.} 
\begin{align}
 \alpha(a, b, c) = & \, \frac{1}{2 \pi} \arctan \sqrt{\frac{b(a+b+c)}{a c}} \\
 =  & \, \frac{1}{2 \pi} \left[ \arctan \left( - 
\sqrt{\frac{a c}{b(a+b+c)}}\right) + \frac{\pi}{2}\right],
\label{eq:risalfa2}
\end{align}
which leads to
\begin{align}
G(\eta) &= \left\{ 
\begin{array}{ll}
 \frac{2}{\pi} \arctan \left( \frac{|1-6\eta|}{\sqrt{12 \eta(1-3 \eta)}} \right) & \mathrm{if} \ \ \eta < \frac{1}{6} \\[5mm]
 \frac{2}{\pi} \arctan \left( - \frac{|1-6\eta|}{\sqrt{12 \eta(1-3 \eta)}} \right) & \mathrm{if} \ \ \eta > \frac{1}{6} 
\end{array} \right. \nonumber \\
& =\frac{2}{\pi} \arctan \left( \frac{1-6\eta}{\sqrt{12 \eta(1-3 \eta)}} \right).
\end{align}
Or, as a function of $\tau$,
 \begin{equation}
  G_{\rm pbc}(\tau) = \frac{2}{\pi} \arctan \left( \frac{1-6\tau(1-\tau)}{\sqrt{12 \tau(1-\tau)(1-3 \tau(1-\tau ))}} \right).
\label{eq:Gdim1pbc}
 \end{equation}
Once more we find an excellent agreement with the numerical data, even
at sizes as small as $L=100$.
This can be seen in Fig.~\ref{fig:dim1pcPer}.

From the analytic expression~\reff{eq:Gdim1pbc}, we see that the
correlation function vanishes when 
\be
1-6 \tau(1-\tau) = 0
\ee 
that is, at
$\overline{\tau}$ and
$1 - \overline{\tau}$, with
\be
\overline{\tau} = \frac{1}{6} \left(3 - \sqrt{3} \right)
= 0.211325\ldots
\ee
which coincides with the numerical value given in~\reff{eq:xbar1dim}.

\section{Conclusions}

We studied the random one-dimensional Euclidean bipartite matching
problem, in which one family of points is on a grid, and the other
family is chosen uniformly at random, when the weight function is the
power $p$ of the Euclidean distance.

At criticality, that is when the two set of points have the same
cardinality, we can solve the problem exactly, for all $p>1$ in the
case of open boundary conditions, and for $p=2$ in the case of
periodic boundary conditions.  Besides these exactly-solvable cases,
other values of the parameter $p$, and the situation in which the two
families of points have different cardinality, have been studied
numerically.

We have computed the average cost and the two-point correlation
function averaged on the distribution of random points. We have
verified by a finite-size scaling analysis the existence of a
nontrivial continuum limit, when the cardinalities of the two sets of
points are equal and sent to infinity. In this limit a close relation
with the Brownian bridge process emerges.  In the exactly soluble
models we find the values $\nu = 2$ and $\alpha =0$ for the critical
exponents which govern the scaling~\reff{scaling}.

It would be interesting to understand the tiny changes in the
two-point correlation function with $p>1$ in the case of periodic
boundary conditions.

\section*{Acknowledgements}

It is a pleasure to thank Massimiliano Gubinelli for very useful
discussions on the continuum limit and Davide Fichera on the Hungarian
Algorithm.

S.C. thanks the Universit\`{e} Paris Nord for the support offered to visit LIPN where this work 
has been finished.

\appendix

\section{Wiener Process and Brownian Bridges}
\label{app}

In this appendix we present some standard notions on Brownian
Processes, which are used in the paper. For a complete source
see for example~\cite{durrett}.

\subsection{Wiener process}
A standard one-dimensional {\em Wiener process}, or {\em Brownian motion process}, is a stochastic process {$W(t)$: $t \in \mathbb{R}, \, t\geq 0$}, with the following properties:
\begin{enumerate}[(1)]
\item $W(0) = 0$
\item The function $t \rightarrow W(t)$ is almost surely continuous 
\item The process $W(t)$ has stationary, independent increments
\item The increment $W(t) - W(s)$ is normally distributed with expected value $0$ and variance $t-s$ 
\end{enumerate}
The requirement that $W(t)$ has {\em independent increments} means
that for all $t_0 < t_1 < \ldots < t_n$, the $n$ random variables
$W(t_1) - W(t_0)$, $W(t_2) - W(t_1)$, $\ldots$, $W(t_n) - W(t_{n-1})$
are independent. The increments are further said to be {\em
  stationary} if, for any $t > s$ and $h > 0$, the distribution of
$W(t+h) - W(s+h)$ is the same as the distribution of $W(t) - W(s)$.

\subsection{Basic properties of the Wiener process}

\begin{itemize}
\item $W(t)$ is a {\em Gaussian process}, that is for all $n$ and times $t_1, \ldots, t_n$,  linear combinations of $W(t_1), \ldots, W(t_n)$ are normally distributed
\item The unconditional probability density function at a fixed time $t$ is given by
\begin{equation}
p_{W(t)}(x) = \frac{1}{\sqrt{2{\pi}t}} e^{-\frac{x^2}{2t}}
\end{equation}
\item $\forall t$, the expectation is zero:
\begin{equation}
{\mathbb E} [W(t)] = 0
\end{equation} 
\item The variance:
\begin{equation}
\var[W(t)]= {\mathbb E} [W^2(t)] - {\mathbb E} ^2[W(t)] = {\mathbb E} [W^2(t)] = t 
\end{equation}
\item The covariance\footnote{To see this, let us suppose $s\leq t$. Then 
\begin{align*}
\cov[W(s), W(t)] &= {\mathbb E} [(W(s)-{\mathbb E} [(W(s)])\cdot (W(t)-{\mathbb E} [(W(t)])] \\&= {\mathbb E} [W(s) \cdot W(t)] = {\mathbb E} [W(s)\cdot((W(t)-W(s))+W(s))]\\
&= {\mathbb E} [W(s)\cdot (W(t)-W(s))]+{\mathbb E} [W^2(s)] = s
\end{align*}}:
\begin{equation}
\cov[W(s), W(t)] = \min(s, t)
\end{equation} 
\end{itemize}
The area of a Gaussian process, defined by
\begin{equation}
W^{(-1)}(t) := \int_0^t \! \mathrm{d} s \, W(s),
\end{equation}
is itself a Gaussian process (as a linear combination of Gaussian processes) characterized by its expected value and variance:
\begin{equation}
{\mathbb E} [W^{(-1)}(t)] = \int_0^t \! \mathrm{d} s \, {\mathbb E} [W(s)] = 0
\end{equation}
\begin{align}
\var[W^{(-1)}(t)] & = {\mathbb E}  \left[ \int_0^t \! \mathrm{d} s \int_0^t \! \mathrm{d} s'  W(s) W(s') \right] 
= \int_0^t \! \mathrm{d} s \int_0^t \! \mathrm{d} s' \cov(W_s, W_s') \nonumber \\
& =  \int_0^t \! \mathrm{d} s \, \left( \int_0^s \! \mathrm{d} s' \, \min(s, s') + \int_s^t \! \mathrm{d} s' \, \min(s, s') \right) = \frac{t^3}{3}.
\end{align}

\subsection{Brownian bridge}
A standard {\em Brownian bridge} $B(t)$ over the interval $[0,1]$ is a standard Wiener process conditioned to have $B(1) = B(0) = 0$.\\
Now, if we have a Wiener process $W(t)$, the linear combination
\begin{equation}
B(t) := W(t) - t\, W(1)
\end{equation}
is a Brownian bridge with expectation, variance and covariance:
\begin{equation}
{\mathbb E} [B(t)] = 0
\end{equation}
\begin{align}
\var[B(t)] & = {\mathbb E} [(W(t) - t\, W(1))^2] \nonumber  \\
&= {\mathbb E} [W^2(t)] - 2 t\, {\mathbb E} [W(1) \cdot W(t)] + t^2\, {\mathbb E} [W^2(1)] \nonumber  \\
& = t\, (1-t)
\label{eq:varBb}
\end{align}
\begin{align}
\cov[B(s), B(t)] &= {\mathbb E} [(W(s) - s\, W(1)) \cdot (W(t) - t\, W(1))]  \nonumber  \\
&= \min(s, t) - s\, t \, .
\end{align}
This covariance is the Green's function of the second derivative with the given boundary conditions. Indeed,
\be
\frac{d^2}{ds^2} \cov[B(s), B(t)]  = - \delta(s-t)
\ee
and this means that the weight of a configuration is
\be
W[B] = \exp \left[ -  \int_0^1 ds \left(\frac{d B}{ds}\right)^2 \right] \, .
\ee
The area of a Brownian bridge, defined by
\begin{equation}
B^{(-1)}(t) := \int_0^t \! \mathrm{d} s \, B(s),
\end{equation}
is, again, a Gaussian variable (as a linear combination of Gaussian random variables) characterized by its expected value and variance:
\begin{equation}
{\mathbb E} [B^{(-1)}(t)] = \int_0^t \! \mathrm{d} s \, {\mathbb E} [B(s)] = 0
\end{equation}
\begin{align}
\var[B^{(-1)}(t)] & = \int_0^t \! \mathrm{d} s  \int_0^t \! \mathrm{d} s' \, \cov[B(s), B(s')] \nonumber \\
& = \int_0^t \! \mathrm{d} s \int_0^t \! \mathrm{d} s' \, \left( \min(s, s') - s\, s' \right) \nonumber \\
& = \frac{t^3}{3} -\frac{t^4}{4}.
\end{align}
In particular, if $t=1$,
\begin{equation}
 \var[B^{(-1)}(1)] = \frac{1}{12}
\end{equation}
and the covariance between $B^{(-1)}(1)$ and $B(t)$ is 
\begin{align}
 \cov[B^{(-1)}(1), B(t)] &= \int_0^1 \! \mathrm{d} s \, \cov[B(s), B(t)] = \int_0^1 \! \mathrm{d} s \, \left( \min(s, t) - s\, t \right) \nonumber \\
&= \frac{1}{2}\, t\, (1 - t).
\end{align}
Let us consider the Brownian Bridge $B$ at two different times, $B(s)$
and $B(t)$, and let us assume that $s \leq t$. The covariance matrix
is then
\begin{equation}
 C = \left( \begin{array}{cc}
            s\, (1-s) & s\, (1-t)\\
            s\, (1-t) & t\, (1-t)
           \end{array}
 \right).
\end{equation}
The distribution is Gaussian, which means the density function is given by
\begin{equation}
p_A(x_1, x_2) = \sqrt{\det A}\ \frac{e^{-\frac{1}{2}\sum_{i=1}^{2}x_iA_{ij}x_j}}{2\pi},
\label{eq:gaussA}
\end{equation}
with
\begin{equation}
  A = C^{-1} = \left( \begin{array}{cc}
            \frac{t}{s(t-s)} & -\frac{1}{t-s}\\
            -\frac{1}{t-s} & \frac{1-s}{(1-t)(t-s)}
\end{array}
 \right)
\end{equation}



\begin{thebibliography}{10}


\bibitem{monge}
G.\ Monge, 
\emph{M\'{e}moire sur la th\'{e}orie des d\'{e}blais et des remblais}, in \emph{Histoire de l'Acad\'{e}mie Royale des Sciences, Ann\'{e}e MDCCLXXXI. Avec les M\'{e}moires de Math\'{e}matiques et de Physique pour la m\^{e}me ann\'{e}e}, Paris, 1784. 

\bibitem{Villani}
C.\ Villani, 
{\em Optimal Transport: Old and New}, Grundlehren der Mathematischen Wissenschaften, Springer, 2008.

\bibitem{kuhn}
H.\ Kuhn, 
{\em The Hungarian Method for the assignment problem}, Naval Research Logistics Quarterly 2, 83--97, 1955.

\bibitem{knuth:sgb}
D.\ E.\ Knuth, \emph{The Stanford GraphBase: A Platform for the Combinatorial computing}, Addison-Wesley, 1993.

\bibitem{munkres}
J.\ Munkres,
\emph{Algorithms for the Assignment and Transportation Problems},  J.\ Soc.\ Ind.\ and Appl.\ Math.\ 5, 32--38, 1957. 

\bibitem{mezard}
M.\ M\'ezard, G.\ Parisi, M.\ A.\ Virasoro, 
\emph{Spin Glass Theory and Beyond}, Word Scientific, Singapore, 1987. 

\bibitem{hartmann}
A.\ K.\ Hartmann, M.\ Weigt, 
\emph{Phase Transitions in Combinatorial Optimization Problems}, John Wiley \& Sons,  2006.

\bibitem{percus}
A.\ Percus, G.\ Istrate, C.\ Moore, 
\emph{Computational Complexity and Statistical Physics}, Oxford University Press,  2006.

\bibitem{bouchaud}
\emph{Complex Systems: Lecture Notes of the Les Houches Summer School 2006}, 
J.-P.\ Bouchaud, M.\ M\'{e}zard, J.\ Dalibard, eds.,
Elsevier,  2007.

\bibitem{montanari}
M.\ M\'ezard, A.\ Montanari, 
\emph{Information, Physics, and Computation}, Oxford University Press,  2009.

\bibitem{MP1}
M.\ M\'ezard, G.\ Parisi,
\emph{Mean-Field Equations for the Matching and the Travelling Salesman Problem},
Europhys.\ Lett.\ 2, 913--918, 1986.

\bibitem{MP2}
M.\ M\'ezard, G.\ Parisi,
\emph{On the solution of the random link matching problem},
J.\ Physique 48, 1451--1459, 1987.

\bibitem{aldous}
D.\ J.\ Aldous,
\emph{The $\zeta (2)$ limit in the random assignment problem}, 
Random Structures and Algorithms 18, 381--418, 2001.

\bibitem{akt}
M.\ Ajtai, J.\ Koml\'{o}s and G.\ Tusn\'{a}dy, 
\emph{On optimal matchings}, Combinatorica 4, 259--264, 1984.

\bibitem{holroyd}
A.\ Holroyd, R.\ Pemantle, Y.\ Peres, O.\ Schramm, 
\emph{Poisson Matching}, Ann.\ Inst.\ Henri Poincar\'{e} Probab.\ Stat.\ 45, 266--287, 2009
(arXiv:0712.1867).

\bibitem{clps}
S.\ Caracciolo, C.\ Lucibello, G.\ Parisi, G.\ Sicuro,
\emph{A Scaling Hypothesis for the Euclidean Bipartite Matching
  Problem}, 
Phys.\ Rev.\ E 90, 012118, 2014,
(arXiv:1402.6993).


\bibitem{MP3}
M.\ M\'ezard, G.\ Parisi,
\emph{The Euclidean matching problem},
J.\ Physique 49, 2019--2015, 1988.

\bibitem{leighton}
T.\ Leighton and P.\ Shor, 
\emph{Tight bounds for minimax grid matching with applications to the average case analysis of algorithms},  Combinatorica 9,  161--187, 1989.

\bibitem{shor}
P.\ W.\ Shor and J.\ E.\ Yukich, 
\emph{Minimax grid matching and empirical measures}, Ann.\ Prob.\ 19, 1338--1348, 1991. 

\bibitem{sicuro}
S.\ Caracciolo and G.\ Sicuro,
\emph{On the one dimensional Euclidean matching problem: exact solutions, correlation functions and universality,}\\
Phys.\ Rev.\ E 90, 042112, 2014,
(arXiv:1406.7565).
 
\bibitem{durrett}
R.\ Durrett, 
\emph{Probability:\ Theory and Examples}, 4th Edition, Cambridge
University Press, 2010. 


\end{thebibliography}
\end{document}